# The Origins of COVID-19

J. C. Phillips

Dept. of Physics and Astronomy, Rutgers University, Piscataway, N. J., 08854

Abstract

The titled subject has attracted much interest. Here we summarize the substantial results obtained by a physical model of protein evolution based on hydropathic domain dynamics.

In a recent Letter eighteen biologists suggested that the titled subject should be studied in a way "inclusive of broad expertise" (*1*). There is an even broader view that has been developed over several decades by physicists (*2,3*). This view is based on analyzing amino acid sequences of proteins. These sequences are now available on-line at Uniprot, and represent a treasure-trove of data (*4*).

Because of large interdisciplinary gaps, the physicists' theory is virtually unknown to biologists. It has been applied to analyze the differences between COVID-03 and COVID-19 (*5*). The special properties here are associated with the spikes, familiar in popular illustrations. The spikes contain ~ 1200 amino acids, and are divided into two major parts, S-1 and S-2. Recent advances in vaccine development have focused on mutations that stabilize S-2 (*6*).

Most of the ~ 300 mutational differences between COVID-03 and COVID -19 are concentrated in S-1. These differences are responsible for the much larger differences in contagiousness that have led to the COVID-19 pandemic. To the author's knowledge, there is only one explanation for the extreme COVID-19 contagiousness, and it can be found only in the phase-transition model (*5*). This model also explains the enhanced contagiousness of recent variants (*7*).

The 300 mutational differences described in (*5,7*) are subtle and complex, and are far too many to have been engineered in a laboratory. They are the result of convergent natural evolution (*8*).